\def\etal{{\it et al.\thinspace}}
\def\eg{{\it e.g.\ }}

\def\ie{{\it i.e.\ }}
\def\cf{{\it c.f.\ }}
\def\gsim{~\rlap{$>$}{\lower 1.0ex\hbox{$\sim$}}}
\def\lsim{~\rlap{$<$}{\lower 1.0ex\hbox{$\sim$}}}

\documentclass[12pt,preprint]{aastex}

\begin{document}

\title{Tides and the Evolution of Planetary Habitability} 

\author{Rory Barnes\altaffilmark{1}, Sean N. Raymond\altaffilmark{2,3}, Brian Jackson\altaffilmark{1}, and Richard Greenberg\altaffilmark{1}}

\altaffiltext{1}{Lunar and Planetary Laboratory, University of Arizona,
Tucson, AZ 85721}
\altaffiltext{2}{NASA Postdoctoral Program Fellow}
\altaffiltext{3}{Center for Astrophysics and Space Astronomy, University of Colorado, Boulder, CO 80309}


\begin{abstract}
Tides raised on a planet by its host star's gravity can reduce a
planet's orbital semi-major axis and eccentricity. This effect is only
relevant for planets orbiting very close to their host stars.  The
habitable zones of low-mass stars are also close-in and tides can
alter the orbits of planets in these locations.  We calculate the
tidal evolution of hypothetical terrestrial planets around low-mass
stars and show that tides can evolve planets past the inner edge of
the habitable zone, sometimes in less than 1 billion years. This
migration requires large eccentricities ($>0.5$) and low-mass stars
($\lsim 0.35 M_\odot$). Such migration may have important implications
for the evolution of the atmosphere, internal
heating and the Gaia hypothesis. Similarly, a planet detected interior to the habitable zone
could have been habitable in the past. We consider the past
habitability of the recently-discovered, $\sim 5$ $M_\oplus$ planet,
Gliese 581 c.  We find that it could have been habitable for
reasonable choices of orbital and physical properties as recently as 2
Gyr ago. However, when we include constraints derived from the
additional companions, we see that most parameter choices that predict
past habitability require the two inner planets of the system to have
crossed their mutual 3:1 mean motion resonance.  As this crossing
would likely have resulted in resonance capture, which is not
observed, we conclude that Gl 581 c was probably never habitable.
\end{abstract}

\clearpage
\section{Introduction}
After the discovery of extra-solar planets, the detection of a
habitable, terrestrial (rocky) planet became a real possibility. Of particular interest are planets in orbit about low-mass stars ($\lsim 0.5 M_\odot$). The lower stellar mass makes detection of lower mass planets easier and leads to habitable orbits with smaller semi-major axes (which also increase the likelihood of detection by current technologies). The
recent detection of putative 5 and 8 $M_\oplus$ planets around an M
star (stellar mass $M_* = 0.31 M_\odot$) (Udry \etal 2007) near the
circumstellar ``habitable zone'' (HZ) (Kasting \etal 1993; Franck \etal
2000; von Bloh \etal 2007; Selsis \etal 2007; hereafter S07) suggests
that Doppler surveys (\ie Butler \etal 2006; Naef \etal 2007) are
becoming sensitive enough to detect terrestrial planets in the
 HZ of low-mass
stars. Not surprisingly, M stars now appear to be attractive targets
for detecting habitable planets (Tarter \etal 2007).

The \textit{habitable zone} is defined, following Kasting \etal
(1993), as the range of distances from a star at which liquid water
could be stable at the surface of a planet. The definition usually
implicitly assumes that the star is the main source of heat. Thus, it
excludes locations where other sources of heat might be available, such
as moons, like Jupiter's Europa, where tidal heating is a major factor
maintaining near-surface liquid water. In that sense, the phrase
\textit{habitable zone}, as usually used in astrophysics, is a
misnomer. Nevertheless, here we use the conventional definition, while
explicitly emphasizing that life could exist outside this zone.

The detection of Gl 581 c and d (Udry \etal 2007) has demonstrated
that planets can form near the HZ of low-mass stars. For
such stars the habitable zone is so close to the star that tidal
effects may be important (Mardling \& Lin 2004). Tides raised between
the star and planet introduce torques on the planet resulting in orbital
evolution, usually decay of semi-major axis, $a$, and eccentricity,
$e$. Here we quantify the effects of tidal forces in such zones. We
find that Earth-sized planets with large initial eccentricities could
migrate out of the HZ of low-mass stars in less than 1
billion years.

The proper definition of an HZ is not obvious. As
described in S07, the boundaries of the HZ are a complex function of
both stellar and planetary properties. For any individual planet, the definitions we choose may not be appropriate. Nonetheless, we will show a general trend that must be
considered when assessing a planet's habitability, both in the past
and in the future.

We consider the evolution of terrestrial planets beginning from a
location that is almost assuredly habitable to a region that is
probably not habitable. We call the time to complete this traversal
the ``habitable lifetime'', $\tau_{HL}$. For relatively massive stars
($\gsim 0.35 M_\odot$), the HZ is far enough away from the star that tides cannot cause significant tidal evolution\footnote{Note we only consider tidal effects after the star has reached the main sequence}.  However for planets with relatively large
eccentricities ($\gsim 0.5$) orbiting low-mass stars ($\lsim 0.35
M_\odot$) tides can pull a planet out of the HZ in less than 1 Gyr. The large number of low-mass stars and significant fraction of
known exoplanets with large $e$ suggest that tides may limit the total
number of habitable planets in the galaxy.

We assume that potentially habitable planets can form with a range of
masses and orbits in the HZ of low-mass stars (see Chambers [2004];
Raymond, Quinn \& Lunine [2007] or Lissauer [2007] for a discussion of
the formation of such planets). Of course, the formation of a planet
in the HZ does not mean a planet is, will be, or has been
habitable. Many other factors are required such as favorable
composition (Lissauer 2007), a thick atmosphere, substantial mass for
sustained plate tectonics ($m \gsim 0.3 M_\oplus$) (Williams, Kasting
\& Wade 1997; Raymond, Scalo \& Meadows 2007), which on Earth drives
the carbonate-silicate cycle (Walker, Hays \& Kasting 1981), and substantial
water content (\eg Raymond, Quinn \& Lunine 2004), both for life and to
facilitate plate tectonics (Regenauer-Lieb, Yuen \& Barlund 2001). Nonetheless we will call planets ``habitable'' if their orbital properties are such that they permit water to be stable on the surface.

The recently announced Gl 581 system contains a planet, c, that is
probably just interior to the HZ (S07). In Table 1 we list the
properties of the planets in this system; $m_p$ is the planet mass, $P$ is
the orbital period, $a$ is the semi-major axis, $e$ is the
eccentricity, $\varpi$ is the longitude of periastron, and $T_{peri}$
is the time of periastron passage in Julian Days. Tidal theory
predicts planet c orbited with larger values of semi-major axis and
eccentricity in the past. Therefore it may have been habitable
in the past, but tides subsequently moved the planet into an
uninhabitable orbit. We find that if planet c was the only planet in
the system, plausible physical properties predict it was
habitable. However, when we consider constraints from the mutual interactions of the
additional planets, we find that planet c has likely never been
habitable.

Tides strong enough to induce significant orbital evolution can also heat the interior of a planet through tidal working. The same type of heating causes Io's intense vulcanism (Peale, Cassen \& Reynolds 1979). Therefore we might expect an increased heating rate on
the planets we consider here as well. The orbital evolution we
consider could also have implications for the evolution of a habitable
planet's atmosphere. In some cases, the orbit-averaged flux can increase
by a factor of 2 as the orbit shrinks (assuming constant stellar luminosity).

In $\S$ 2 we review tidal theory, present our definition of the HZ,
and describe how we determine the habitable lifetime. In $\S$ 3 we
present our results for hypothetical terrestrial planets. In $\S$ 4 we
describe how our model applies to the Gl 581 c planet. Finally, in
$\S$ 5 we draw our conclusions and discuss directions for future
research.

\section{Methodology}

\subsection{Tidal Theory}
We evaluate the effects of tides using the classical, second order
equations assembled by Goldreich \& Soter (1966) (see also Kaula
1964; Greenberg 1977):
\begin{equation}
\label{eq:adot}
\frac{da}{dt} = -\Big(21\frac{\sqrt{GM_*^3}R_p^5k}{m_pQ_p}e^2 + \frac{9}{2}\frac{\sqrt{G/M_*}R_*^5m_p}{Q'_*}\Big)a^{-11/2}
\end{equation}
\begin{equation}
\label{eq:edot}
\frac{de}{dt} = -\Big(\frac{21}{2}\frac{\sqrt{GM_*^3}R_p^5k}{m_pQ_p} + \frac{171}{16}\frac{\sqrt{G/M_*}R_*^5m_p}{Q'_*}\Big)a^{-13/2}e
\end{equation}
where $a$ is semi-major axis, $G$ is Newton's gravitational constant,
$m_p$ is the mass of the planet, $k$ is the planet's Love number,
$Q_p$ is the planet's tidal dissipation function, $Q'_*$ is the star's
tidal dissipation function divided by two-thirds its love number,
$R_p$ is the planet's radius, and $R_*$ is the star's radius. The
first terms in these equations are due to the tide raised on the
planet by the star, and the second terms are due to tides raised on
the star by the planet. Eqs.\ (\ref{eq:adot} -- \ref{eq:edot}) are
only valid if the planet's orbital period is less than the star's
rotation period, otherwise the signs of the second two terms will
change. For the cases we consider, this condition is irrelevant
because the second terms of Eqs.\ (\ref{eq:adot} -- \ref{eq:edot}) are
6 -- 7 orders of magnitude smaller than the first. This difference is
largely due to the mass difference between a terrestrial planet and a
star. Note that tidal theory predicts decay for both $a$ and $e$. As
tides slowly change a planet's orbit, the planet may move into or out
of the HZ.

The equations above were derived under the implicit assumption that $e$ 
values are small. Here we apply them to cases that include conditions
with large $e$ values. Thus higher-order corrections may potentially be
important. Several models of tidal evolution that are not restricted to
small values of e have been derived (Hut 1981; Eggleton, Kiseleva \& Hut
1998; Mardling \& Lin 2002). However, such models must include specific
assumptions about how a body responds to the ever-changing tidal
potential. The tidal forcing involves components with a various
frequencies, but the nature of the response of a planet or star is
uncertain. It is not clear how the tidal phase lag depends on the
frequency of each harmonic. This issue becomes increasingly critical when
$e$ is large and a wide range of different comparable-amplitude frequencies
are involved. For a model valid to a high order in $e$, the validity of
parameterizing tidal effects through a single constant $Q$ is questionable.
Moreover, tidal heating could lead to core/mantle melting and a time-dependent $Q$ value. Not enough is known about the actual response of real bodies to evaluate
these higher-order effects. Therefore, to the best of current knowledge,
Eqs.\ (1 -- 2) above are reasonable representations of tidal evolution. As
the theory matures, the effect of higher-order terms could be included
and this study should be revisited. Note that the higher order theories
that have been derived all predict faster evolution than Eqs. (1 -- 2)
(all the higher order terms have the same sign as those in Eqs. (1 -- 2), so their predictions for the habitable lifetime, $\tau_{HL}$, would
be even shorter than those we derive below. In that sense, our approach
provides a conservative estimate of the tidal effects on terrestrial
planets.

Eqs.\ (\ref{eq:adot} -- \ref{eq:edot}) are only valid when the star's
orbit-averaged torque on the planet's tidal bulge is zero. When this situation occurs, the planet is said to be ``tidally locked'' to its star. For a
planet on an eccentric orbit, the torque is greatest at pericenter at
which point the star is moving with an angular velocity greater than
the mean motion. Therefore the rotational frequency of the planet
$\Omega$ is larger than its orbital frequency $n$ (tidal locking does not mean that one side of the planet faces the star for all times, this situation only occurs for a circular orbit). To second order in $e$ these two
frequencies are related by the following equation,
\begin{equation}
\label{eq:spinlock}
\Omega = n(1 + \frac{19}{2}e^2)
\end{equation}
(Murray \& Dermott 1999; Goldreich 1966). Note that this equation neglects effects of a permanent quadrupole moment (we ignore this effect as their is currently no hope of measuring it in an exoplanet). Even for modest eccentricities, the spin and orbital frequencies can be significantly different due to the coefficient in front of the $e^2$ term. A planet
with an eccentricity of 0.32 has a spin period half that of its
orbital period.

The time for a planet to become tidally locked is given by
\begin{equation}
\label{eq:timelock}
t_{lock} = \frac{8m_pQ_pa^6}{45GM_*^2kr_p^3}\Delta\Omega
\end{equation}
(Goldreich \& Soter 1966; Rasio \etal 1996), where $\Delta\Omega$ is the difference between the initial spin, and the spin given by Eq.\ (\ref{eq:spinlock}). For an Earth-mass, terrestrial planet ($k=0.3, Q_p = 100$) planet
orbiting a 0.3 $M_\odot$ star with $e=0$ $a=0.1$ AU, this timescale is about $10^6$ years. We
consider planets initially with much larger eccentricities, but
eccentricity probably does not affect this timescale significantly
(Peale 1977). This time is very short compared to other timescales associated with planet formation, and we may assume
the planets are tidally locked to the star as soon as they are formed.

In order to determine $\tau_{HL}$ for possible planets, we consider a
range of values of $M_*$, $m_p$, $a$ and $e$. We will set $Q'_* =
10^{5.5}$, $Q_p = 21.5$, and $k = 0.3$, reasonable choices based on
limited observational constraints for stars (Ogilvie \& Lin
2007; Jackson, Greenberg \& Barnes 2007) and the Earth (Dickey \etal 1994; Mardling \& Lin 2004).  However, the
time-averaged value of $Q$ of the Earth is
probably closer to 100 (Lambeck 1977), which is similar to the values
of other terrestrial planets in the Solar System (Yoder
1995). Nonetheless we choose $Q_p = 21.5$ to be consistent with past
work (Mardling \& Lin 2004).

From
our choices of $M_*$ and $m_p$ we must determine radii. For the
planet, we will use the scaling relationship
\begin{equation}
\label{eq:rp}
\frac{R_p}{R_\oplus} = \Big(\frac{m_p}{M_\oplus}\Big)^{0.27}
\end{equation}
(Valencia, O'Connell \& Sasselov 2006). For the stellar radius, we use an empirical
relation derived from observations of eclipsing binaries (Gorda \&
Svechnikov 1999):
\begin{equation}
\label{eq:rs}
\log_{10} \frac{R_*}{R_\odot} = 1.03\log_{10} \frac{M_*}{M_\odot} + 0.1,
\end{equation}  
where $R_\odot$ and $M_\odot$ are the solar radius and mass,
respectively. Note that Eq.\ (\ref{eq:rs}) is only valid when $M_*
\lsim M_\odot$.

\subsection{The Eccentric Habitable Zone}
For circular orbits, the HZ is defined as the range of orbits in which
a terrestrial planet, with an atmosphere favorable for habitability, can support liquid
water on the surface (Kasting \etal 1993; S07). However, extra-solar
planet orbits are often highly eccentric. For these orbits,
Williams \& Pollard (2002) showed that, despite large variations in
surface temperature, the key for long-term climate stability is the
orbit-averaged flux,
\begin{equation}
\label{eq:flux}
F = \frac{L}{4\pi a^2\sqrt{1-e^2}}, 
\end{equation}
(Williams \& Pollard 2002; Adams \& Laughlin 2006), where $F$ is the orbit-averaged flux and $L$ is the stellar luminosity. Note, however, that potential water loss at
periastron and potential cloud freeze-out at apoastron have not been
accounted for, despite their implications for habitability. We define
the eccentric habitable zone (EHZ) as the range of orbits for which a
planet receives as much flux over 1 orbit as a planet on a circular
orbit in the ``classical'' HZ (Kasting \etal 1993; S07). We must also scale the
location of the EHZ to the luminosity of the stars we are
considering. Lower mass stars have lower luminosities, and hence the
EHZ's are closer to the star. We therefore define the inner and outer
edges of the EHZ, $l_{in}$ and $l_{out}$ (in AU), respectively, as
\begin{equation}
\label{eq:lin}
l_{in} = (l_{in\odot} - a_{in}T_* - b_{in}T_*^2)\Big(\frac{L}{L_\odot}\Big)^{1/2
}(1-e^2)^{-1/4},
\end{equation}
and
\begin{equation}
\label{eq:lout}
l_{out} = (l_{out\odot} - a_{out}T_* - b_{out}T_*^2)\Big(\frac{L}{L_\odot}\Big)^{1/2
}(1-e^2)^{-1/4},
\end{equation}
where $l_{in}$ and $l_{out}$ are the inner and outer edges of the HZ, respectively, in AU, $l_{in\odot}$ and $l_{out\odot}$ are the
inner and outer edges of the HZ in the solar system, respectively, in AU, $a_{in} = 2.7619
\times 10^{-5}$ AU/K, $b_{in} = 3.8095 \times 10^{-9}$ AU/K$^2$, $a_{out} = 1.3786 \times 10^{-4}$ AU/K, and $b_{out} = 1.4286 \times 10^{-9}$ AU/K$^2$ are empirically
determined constants, $T_* = T_{eff} - 5700$ K, and $L_\odot$ is the solar luminosity (S07). Terms involving $T_*$ relate stellar intensity at different wavelengths and atmospheric windows of predicted habitable planets.  Eq.\ (\ref{eq:lin} -- \ref{eq:lout}) are only valid for $T_{eff} < 3700$ K. $T_{eff}$ is the
effective temperature of the star, given by
\begin{equation}
\label{eq:teff}
T_{eff} = \Big(\frac{L}{4\pi\sigma R_*^2}\Big)^{1/4}
\end{equation}
where $\sigma$ is the Stefan-Boltzman constant.Therefore a planet with
a given $e$ is on the inner edge of the EHZ if $a = l_{in}$. Note that as $e$ changes so do the locations of $l_{in}$
and $l_{out}$.

The values of $l_{in\odot}$ and $l_{out\odot}$ are therefore the key
parameters in the identification of the edges of the EHZ. We
consider three criteria identified in S07: 1) 0\% cloud cover, 2) 50\%
cloud cover, and 3) 100\% cloud cover\footnote{Note that S07 also give
values of $l_{in\odot}$ and $l_{out\odot}$ based on assumptions
regarding the past and present habitability of Venus and Mars coupled
with stellar evolution, but we will ignore those models here.}. S07
give values of $l_{in\odot}$ of $\sim 0.89$, $\sim 0.72$, and
$\sim 0.49$ AU for the three possibilities, respectively. For the outer
edge, S07 give $l_{out\odot} = 1.67, 1.95,$ and 2.4 AU, for the three cloud cover models, respectively. We
arbitrarily choose the 50\% cloud cover case to be the limits of the
EHZ. This choice is the middle of the possibilities, and roughly
corresponds to the cloud cover on the Earth.

The stellar mass is related to the luminosity in the following way,
\begin{equation}
\label{eq:luminosity}
\lambda = 4.101\mu^3 + 8.162\mu^2 + 7.108\mu + 0.065,
\end{equation}
where $\lambda$ = log$_{10} (L/L_\odot)$ and $\mu$ = log$_{10}
(M_*/M_\odot)$ (Scalo \etal 2007). With these definitions, we plot the shape of the EHZ
as a function of stellar mass and initial eccentricity, $e_0$, in
Fig.\ \ref{fig:hz}. As both $e_0$ and $M_*$ increase, the location of
the EHZ moves out and grows wider. If, for a certain star, $T_{eff} < 3700$ K, we set it to 3700 K in Eqs.\ (\ref{eq:lin} -- \ref{eq:lout}) as done in S07.

\subsection{Determination of the Habitable Lifetime}
The habitable lifetime, $\tau_{HL}$, depends on the initial orbit of a
planet. Rather than explore all possible orbits, we will consider one
type of starting location: the inner 0\% cloud cover boundary (\ie
$l_{in\odot} = 0.89$ AU). Although this choice
of initial condition is arbitrary, it demonstrates how tides will
change the orbits, and, hence, habitability. As stated in the previous
section, we consider a planet uninhabitable when it crosses the 50\%
cloud cover boundary ($l_{in\odot} = 0.72$ AU). We vary the planet's initial eccentricity $e_0$ and $M_*$ in
increments of 0.01 and 0.01 $M_\odot$, respectively, and consider four different planet masses: 5, 1,
0.5, and 0.1 M$_\oplus$.

In order to calculate $\tau_{HL}$ we numerically integrate Eqs.\
(\ref{eq:adot} - \ref{eq:edot}) and (\ref{eq:lin}) with a timestep of
10,000 years. Convergence tests showed that this timestep is two
orders of magnitude smaller than necessary to produce reliable
results. We integrate the tidal equation until either $a < l_{in}$(50\% cloud cover) or
$\tau_{HL} > 10$ Gyr.

\section{Habitable Lifetimes}
In this section we present results for $\tau_{HL}$ as defined in $\S$
2.3. Fig.\ \ref{fig:thz4} shows contour lines of $\tau_{HL}$ for planets
with masses of 5, 1, 0.5 and 0.1 M$_\oplus$. There are two prominent
features in each panel. First is the abrupt change in $\tau_{HL}$ from
10 Gyr to less than 1 Gyr between $e_0 = 0.53$ and 0.54. This feature
occurs because the time to circularize the orbit is less than 10 Gyr. When $e$ reaches zero, the
evolution of $a$ effectively stops, and the planet is stuck on one
side of the boundary or the other. Orbits with $e_0 \lsim 0.53$ become
circular with $a$ just larger than $l_{in}$. However, when $e_0 \gsim
0.54$, the orbits circularize with $a$ slightly less than $l_{in}$.

These different types of evolution are shown in Fig.\ \ref{fig:evol}.
Note that the evolution of both cases is very similar, but the final
relationship between $a$ and $l_{in}$ is such that one planet remains
in the EHZ, but the other does not. Note that the EHZ changes because
$e$ is changing. The difference is therefore not due to some fundamental
feature of tidal theory (other than changes in semi-major axis become
negligible for small eccentricities), rather
our definition of the EHZ leads to the different values of
$\tau_{HL}$.

The apparently sudden transition from long to short habitable
lifetimes can be shown by considering how $a$ and $e$ are related:
\begin{equation}
\label{eq:ae}
\ln\Big(\frac{a}{a_0}\Big) = e^2 - e_0^2,
\end{equation}
where $a_0$ and $e_0$ are the initial values of $a$ and $e$, if the effects of the tide raised on the star by the planet is
negligible (which is the case here) (Jackson, Greenberg \& Barnes 2007). If we assume $e=0$ (\ie the tidal evolution has
effectively ended, and $t = \infty$), $a = l_{in}$(50\% cloud cover)
and $a_0 = l_{in}$(0\% cloud cover), then the value of $e_0$ that
solves Eq.\ (\ref{eq:ae}) should correspond to the critical value ($e_0 \sim 0.535$)
seen in Fig.\
\ref{fig:thz4} (note that $a_0$ is also a function of $e_0$ and Eq.\
[\ref{eq:ae}] must be solved numerically). In the cases we consider,
this equation is independent of mass. The values of $a$ and $a_0$ have
the same dependence on luminosity (and hence mass, \cf Eq.\
[\ref{eq:luminosity}]), and, for the stars we consider, we have set
$T_{eff} = 3700$ in order to follow the example of S07. Therefore we
expect the critical value of $e_0$ to be independent of both stellar
and planetary mass. Solving Eq.\ (\ref{eq:ae}) for the values of
$l_{in}$ we have chosen, we find the critical value of $e_0$ is
0.53490, in agreement with Fig.\ \ref{fig:thz4}.

The second feature of the contours is the downturn that occurs for
$e_0 \gsim 0.65$, in other words, $\tau_{HL}$ gradually increases with
$e_0$. This dependence follows from the change in the EHZ at larger
values of $e_0$, as shown in Fig.\ \ref{fig:hz}. At these values not
only is the EHZ wider, but $a_0$ is also larger. Therefore planets' orbits
will evolve more slowly (note the $a$ dependence in Eqs.\ [\ref{eq:adot} --
\ref{eq:edot}]), and have further to go before they reach $l_{in}$(50\% cloud cover).

We also see in Fig.\ \ref{fig:thz4} that lower mass planets tend to
have longer habitable lifetimes. This relationship follows from the
dependence of $da/dt$ on $m_p$ in Eqs. (\ref{eq:adot} --
\ref{eq:edot}). Low-mass stars, like the ones considered here, are
likely to host low-mass planets (Laughlin, Bodenheimer \& Adams 2004; Ida \& Lin 2005; Raymond, Scalo \& Meadows 2007), and therefore disks around M
stars may preferentially form planets with relatively long habitable
lifetimes. However, they may not contain enough volatiles for life
because the impacts that deliver them may have large enough energies that the volatiles will be lost to the planet (Lissauer 2007).

Models of the evolution of habitable atmospheres have focused on planets with 
static orbits (\eg Kasting \& Catling 2003). Figure \ref{fig:thz4}
shows that such a model may be inappropriate for determining
habitability and biosignatures of planets around M stars (Segura \etal
2005; S07). In Fig.\ \ref{fig:flux} we plot the orbit-averaged stellar
flux for a planet with
$\tau_{HL} = 4.5$ Gyr orbiting a 0.2 $M_\odot$ star (assuming the
stellar luminosity is constant). This figure also shows how the spin
frequency of the planet $\Omega$ evolves relative to its mean motion $n$,
see Eq.\ (\ref{eq:spinlock}). The tidally evolving planet does not
orbit such that only one side faces the star until the orbit has
become circular at $t \approx 6$ Gyr.

Note as well that there exists a family of orbits with $\tau_{HL}
\approx 4.5$ Gyr, about the age of the Earth. Therefore these
calculations suggest that if a planet with the correct ingredients for
life formed with that orbit, and life subsequently developed in a
manner similar to the Earth, that life would be eliminated by the
ultimate global warming event: The passage of the planet through the
inner edge of the HZ due to tidal evolution.

\section{Application to the Gl 581 System}
In this section we apply the ideas of $\S$ 2 to the recently
announced Gl 581 planetary system (Udry \etal 2007). This system
contains three planets with orbits and masses listed in Table
1. Planet c is probably interior to the 50\% cloud cover HZ, and
planet d is probably exterior (S07). The age of the star is
uncertain. Its kinematics and metallicity suggest that it is at least a few Gyr old (Bonfils \etal 2005; Delfosse \etal 1998), but its very low X-ray luminosity (Delfosse \etal 1998) indicates it may be as old as 8 --
10 Gyr old (Ribas, private communication; Selsis, private
communication). We therefore tidally evolve planet c backward for 10 Gyr by
changing the signs of Eqs.\ (\ref{eq:adot} -- \ref{eq:edot}) (Jackson, Greenberg \& Barnes 2007). First we consider how Gl 581 c would evolve if
it were an isolated planet, then we consider how the additional
companions constrain its evolution and the planets' physical
properties.

\subsection{Past Evolution of Gl 581 c}
If Gl 581 c were the only planetary companion to its host star, then
its past tidal evolution is adequately modeled by Eqs.\ (\ref{eq:adot} --
\ref{eq:edot}). For this particular system, we have specific
constraints on physical and orbital properties. The planet's radius
$R_c$ lies in the range $1.6 \lsim R_c \lsim 2 R_\oplus$, assuming the
observed minimum mass is the actual mass (S07; Valencia, O'Connell \& Sasselov 2006;  Fortney, Marley \& Barnes 2007; Valencia, Sasselov \& O'Connell 2007a, 2007b; Sotin, Grasset \& Moquet 2007), and its best-fit eccentricity
$e_c$ is 0.16, although any value between 0 and 0.3 is about equally likely
(Udry \etal 2007). For tidal parameters, we appeal to values in our
Solar System. The current values of $k$ and $Q$ for the Earth are
about 0.3 and 21.5, respectively, from Lunar Laser Ranging (Dickey
\etal 1994; Mardling \& Lin 2004), but over the lifetime of the
Earth-moon system, the $Q$ value is probably close to 100 (Lambeck
1977), similar to the value of Mars (Yoder 1995). The
star's mass, radius, luminosity and effective temperature are $0.31
M_\odot, 0.38 R_\odot, 0.0135 L_\odot$ and 3200 K, respectively
(Bonfils \etal 2005). Planet c is currently inside the 100\% cloud
cover HZ, but such an atmospheric condition is unlikely (S07). We
therefore seek to identify physical and orbital parameters that allow Gl 581 c to have been habitable in the past, \ie inside the 50\% cloud cover HZ.

First we consider the effects of varying eccentricity. In Fig.\
\ref{fig:e} we show how long ago Gl 581 c would have been inside the
50\% cloud cover HZ, as a function of its current eccentricity, for 2
different, plausible values of $Q_c$, assuming $k = 0.3$. At low
$e_c$ values, the time is longer because tidal evolution is slower
with smaller $e$, while at larger values the time is longer due to the strong coupling between $a$ and $e$, and because the
edge of the HZ is farther out, \cf Eqs.\ (\ref{eq:adot} -- \ref{eq:edot}) 
and (\ref{eq:lin}). Therefore
there exists a critical value of $e_c$ that minimizes the time since the
planet was habitable. In this case, that value is 0.32 in which case
the planet would have been inside the 50\% cloud cover HZ as recently
as 3 Gyr ago.

Next we consider how the planet's radius constrains past
evolution. For most cases we considered, the Gl 581 c planet could
have been inside the 50\% cloud cover HZ for many Gyr, assuming a
system age of 10 Gyr. In Fig.\ \ref{fig:rp} we plot the evolution of
the EHZ and $a_c$ for three different values of $R_p$, and assuming
its current eccentricity is 0.16. All the cases plotted assume $Q_c =
21.5, k = 0.3$. The shading represents the 100\%, 50\%, and 0\% cloud
cover models, and white is outside the HZ, as in Fig.\
\ref{fig:hz}. Even for the extreme case $r_c = 2 R_\oplus$, the planet
could never have been in the 0\% cloud cover HZ. In this case, it is
hard to imagine how $Q_c$ could be so low, since the planet is most
likely a ``water world'' (Raymond, Quinn \& Lunine 2004; L\'eger \etal
2004) and for such a planet dissipation of tidal energy would be less
efficient ($Q$ would be larger). If the system
is 10 Gyr old, this case predicts planet c was habitable for 8 Gyr and
tides would have sterilized planet c 2 Gyr ago.

\subsection{Possible Interactions Between Planets b and c}
So far we have ignored interactions between planets b and c, but as we run
backwards in time, we need to remember that their mutual gravitational interactions will cause the eccentricities to oscillate on a timescale of $\sim 10^3$ years, and we must consider the possibility that as their
periods change, they could hit a mean motion resonance.  The first
significant resonance would be the 3:1 ratio of orbital periods. If
they ever crossed this resonance, the planets would have almost
assuredly been captured into the resonance (Lee \& Peale 2002). Such a
scenario would invalidate the conclusions of the previous section because the system is not observed to be in resonance today. 

Planet b is also subject to tidal forces so we must account for its
tidal evolution. Currently its outer 3:1 resonance lies at $\sim
0.085$ AU, and therefore if it did not experience significant
evolution, then planet c has orbited inside that distance since the
system formed. If, however, planet b experienced significant tidal
evolution, then both planets could have migrated such that the 3:1 was
always instantaneously beyond the orbit of planet c. The value of
$Q'_b$ is unknown for the Neptune-mass planet. If the planet is
gaseous ($k = 3/2$), then $Q_b \sim 10^6$ (Jackson, Greenberg \& Barnes
2007), but if it is terrestrial, then the value is probably closer to
100. In Fig.\ \ref{fig:qb} we plot the evolution of $a_b$ and $e_b$
for values of $Q_b$ between $10^4$ and $10^6$. We assume the planet
has the same bulk density as Neptune, and therefore the radius of
$3.74 R_\oplus$. This choice is somewhat arbitrary, but is consistent
with the radius of the Neptune-mass planet GJ 436 (Gillon \etal 2007;
Deming \etal 2007). The evolution depends strongly on $Q_b$, and in
order for planet c to have been habitable and avoided the 3:1
resonance, then $Q_b \lsim 5\times 10^4$.

In fact, $Q_b$ must be less than $4\times 10^4$ in order to avoid the
3:1 resonance, as shown in Fig.\ \ref{fig:3_1}. This figure plots the
evolutions of b and c (without mutual interactions, and $Q_c = 21.5,
r_c = 1.8 R_\oplus$), and the instantaneous 3:1 resonance of planet
b. We see that in this case planet c just misses the 3:1 resonance.

To this point, we have not considered the secular interactions between
the planets. Secular interactions cause the eccentricities and
orientations of the orbits to oscillate with time (see \eg Barnes \&
Greenberg [2006] for a review of secular theory). In Fig.\
\ref{fig:secular} we show the secular interactions for the currently
observed system, excluding planet d. This simulation was performed
with HNBODY\footnote{Publicly available at
http://janus.astro.umd.edu/HNBody}, and includes general
relativistic effects. Our integration reveals that the current
values of the eccentricities are near their extrema. We presume that
using the observed values provides a reasonable estimate of the tidal
evolution, however we may be overestimating the tidal effects on
planet c, and underestimating the tidal effects on planet b. A more sophisticated treatment that combines the secular and tidal evolutions was beyond the scope of this investigation (see $\S$ 5).

Tidal theory predicts the eccentricities would be larger in the
past. The amplitude of the eccentricity oscillations (due to gravitational interactions between planets) scales with $e$
(among other quantities), and eccentricity oscillations would
therefore have been larger in the past. Larger eccentricity values
can often lead to dynamical instability (\eg Barnes \& Quinn 2004;
Barnes \& Greenberg 2007). Therefore we also require that the
mutual, tidal evolutions of planets b and c predict dynamically stable orbits. Such a consideration shows that if $Q_c$ is small
enough to exclude a 3:1 resonance capture (\ie $< 4 \times 10^4$),
then the orbits of planets b and c would have been unstable 10 Gyr
ago. This instability arises from the large $e$ values the planets had
then ($e_b = 0.74, e_c = 0.47$).

Even if we begin the backward integration with the two planet's
average eccentricity (over the secular cycle), planet c was probably
not habitable. Planet c's average eccentricity is lower than its
observed value, and it has probably experienced less tidal evolution than
predicted above. Therefore it is unlikely it could have ever been
inside the 50\% cloud cover EHZ. On the other hand, b's average
eccentricity is larger than the current value, and its tidal evolution
would have probably been more significant. In that case, tidal evolution is
more likely to predict dynamical instability. We conclude that the
requirements that planets b and c avoid the 3:1 resonance and be
stable at all times cannot be satisfied if $Q_c = 21.5$ and $k=0.3$.

However, if the two planets have properties similar to analogous bodies in the
Solar System, then the system would have been stable, and never in a
3:1 resonance. In Fig.\
\ref{fig:bestrewind} we show the evolution of the 2 planets assuming
the inner planet is Neptune-like ($Q_b = 10^6, R_b = 3.74 R_\oplus$),
and planet c is a typical terrestrial planet ($Q_c = 100, k = 0.3, R_c =
1.8 R_\oplus$) (Yoder 1995). In this case, the 3:1 resonance crossing
occurred nearly 10 Gyr ago, and the system 10 Gyr ago is stable
(assuming low amplitude apsidal libration). This scenario invokes the
simplest assumptions, and is therefore the
preferable explanation. We conclude that Gl 581 c is probably a terrestrial planet with $Q \sim 100$, but was never inside the
50\% cloud cover EHZ.

\section{Conclusions}
The detection of a terrestrial planet around a low-mass star is insufficient to 
determine that planet's past and future habitability. The tidal forces
between planet and star can significantly change the orbits, and hence
limit the habitable lifetime. Planets detected in the HZ with large
eccentricities may be bound for hotter temperatures and ultimately a
global extinction. On the other hand, planets detected interior to the HZ
may have been habitable in the past. Gl 581 c was most likely not
habitable in the past, but $\S$ 4.1 shows that if its companion planets
were on different orbits, past habitability would have been possible.

For plausible physical and orbital properties of hypothetical
terrestrial planets, tides may evolve the planets' orbits past the
inner edge of the HZ over a timescale comparable to the age of the
Earth. Therefore in order for planets to be habitable long enough for complex
life to develop, they must form with eccentricities low enough that
tides don't eventually make them inhospitable to life. Alternatively
planets that form beyond $l_{out}$ may evolve into the habitable zone,
but the ``flux'' of planets out, through the inner edge of the EHZ, is larger
than that of planets in, through the outer edge, because of the very steep
dependence of $da/dt$ on $a$. Therefore, on average and assuming
uniform distribution in $a$, tides tend to reduce the total number of habitable
planets around M stars in the galaxy.

Our work has shown that tides may shorten habitable lifetimes of
planets that form with eccentricities larger than about 0.5. Although
the formation mechanism of such planets is unknown, the existence of
giant planets with such eccentricities suggests even terrestrial-sized
planets may form with similar values. However, the majority of known
exoplanets probably formed with eccentricities below 0.5 (see, \eg Jackson,
Greenberg \& Barnes 2007), thus most habitable lifetimes
will not be significantly shortened due to tides.

This work predicts that the vast majority of terrestrial planets in
HZs around low-mass stars ($\lsim 0.2 M_\odot$) will be detected on nearly
circular orbits because the time to circularize the orbit is less than 1 Gyr, \cf Fig.\ \ref{fig:evol}. Should space-based transit
missions like COROT and Kepler find terrestrial planets on non-circular orbits around old
(more than a few Gyr), low-mass stars ($< 0.35 M_*$) in small orbits
($a\lsim 0.1$ AU), then that system is likely to contain additional companions (Mardling \& Lin 2004) which pump up eccentricities due to mutual interactions.

The increasing stellar flux on a planet due to orbital decay could
impact its atmospheric conditions and composition. Previous models of
habitable atmospheres have considered planets at a single semi-major
axis value (\ie Kasting \etal 1993; Kasting \& Catling 2003; Segura
\etal 2005; Tinetti, Rashby \& Yung 2006; Kiang \etal 2007). The tidal change
in the orbit of a habitable planet may affect the relative abundances
of biosignatures, such as the simultaneous presence of oxygen and
methane (Sagan \etal 1993). Future work on the evolution of a
habitable atmosphere and the identification of biosignatures should
consider the change in stellar flux due to tidal evolution.

When a planet is detected such that it was most likely in the HZ in
the past, it would be interesting to determine if it did, in fact, support life in the past. In the foreseeable future,
such a determination will have to be made through analysis of disk-averaged
spectra of planets (Tinetti \etal 2006; Tinetti, Rashby \& Yung 2006; Kaltenegger, Traub \& Jucks 2007). Future work on the evolution of habitable atmospheres should
explore the possibility of detecting signatures of extinct life on
planets around M stars.

Future work should also address the determination of habitable lifetimes in multiple
planet systems as well as incorporate higher
order corrections to the tidal equations. Such improvements will
require substantial advancement in our understanding of the
deformations of solid and gaseous bodies due to tides. The Gl 581
systems demonstrates the need to develop a theory that describes two
tidally damped orbits that also experience significant mutual
interactions (note that Mardling [2007] has derived expressions for systems in which 1 planet experiences tidal evolution). However, the results of $\S$ 4.2 suggest that such improvements are unlikely to change our
assessment for the Gl 581 system.

Planets with masses $\lsim 0.3
M_\oplus$ are likely too small to support plate tectonics, which are
thought to be necessary for life (Williams, Kasting \& Wade 1997; Raymond, Scalo \& Meadows 2007). This work suggests some of these smaller planets may be heated
internally by tides, such as in the Galilean satellites (Peale, Cassen \& Reynolds 1979). In some cases, in which $\tau_{HL}$ is long
enough for life to develop, the tidal working of the body may supply
enough heat to drive plate tectonics after radiogenic heating has become negligible. Therefore these low-mass planets, previously thought to be
uninhabitable, may, in fact, be good locations for life. Future work should
explore this possibility.

Perhaps the most distressing aspect of this work (from a SETI
perspective) is the prediction that planets can be habitable long
enough for complex life to develop, but then that life is extinguished
by tides. Yet this work suggests that such a ``tidal extinction'' may
occur on some planets around low-mass stars.

The results presented here may have significant implications in the
context of the Gaia hypothesis (Lovelock \& Margulis 1974), which
proposes that biology can change the physical conditions on a planet,
such as its atmosphere, so as to promote the continuing existence of
life. In other words, evolution favors organisms that help maintain a
habitable environment. According to that hypothesis, an inhabited
planet may avert a tidal extinction due to its organisms developing
adaptations that not only change themselves, but also change the atmosphere such that
the sterilizing effects of increased stellar radiation are
mitigated. Therefore the Gaia hypothesis predicts that we may observe
biosignatures on planets that are on orbits interior to the nominal
HZ, if they were formerly in the HZ. In that case we might need to
refine the definition of the HZ into two branches: the ``physical'' HZ
(\ie the definition proposed by Kasting, Whitmire \& Reynolds [1993]
that applies to planets before life gains a foothold), and the
``biologically-extended'' HZ (\ie a larger region in which evolved
life can stave off sterilization). For example, planet Gl 581 c was
probably never in the physical HZ, but if it once had been, then it
might have remained habitable even after reaching its current orbit,
according to the Gaia hypothesis. The detection of other similar new
planets may provide observational tests of the Gaia hypothesis. The
detection of such a planet would be a landmark event, not only in the
detection of life beyond the Solar System, but because it will reveal
fundamental features of life, evolution, and planetary habitability.

We conclude that tides can move habitable planets into non-habitable
orbits and described how this phenomenon is relevant for astrobiological models.  Future work into the search for habitable worlds, through
planet-hunting, atmospheric modeling, geophysical modeling, etc.,
should bear in mind the effects of tidal evolution of planets around M
stars.

\medskip
This work was funded by PG\&G grant NNG05GH65G. S.N.R. was supported by an appointment to the NASA Postdoctoral Program t the University of Colorado Astrobiology Center, administered by Oak Ridge Associated Universities through a contract with NASA.

\references
Adams, F.C. \& Laughlin, G. (2006) Effects of Secular Interactions in Extrasolar Planetary Systems. \textit{Astrophys. J.} 649, 992-1003.\\
Barnes, R. \& Greenberg, R. (2006) Extrasolar Planetary Systems Near a Secular Separatrix. \textit{Astrophys. J.} 638, 478-487.\\
---------------. (2007) Stability Limits in Resonant Planetary Systems. \textit{Astrophys. J. Lett.}, 665, L67-L70.\\
Barnes, R. \& Quinn, T.R. (2004) The (In)stability of Planetary Systems. \textit{Astrophys. J.} 611, 494-516.\\
Bonfils, X. \etal (2005) The HARPS search for southern extra-solar planets. VI. A Neptune-mass planet around the nearby M dwarf Gl 581. \textit{Astron. \& Astrophys. Lett.} 443, L15-L18.\\
Butler, R.P. \etal (2006) Catalog of Nearby Exoplanets. \textit{Astrophys. J.} 646, 505-522.\\
Chambers, J.E. (2004) Planetary accretion in the inner Solar System. \textit{Earth \& Plan. Sc. Lett.} 223, 241-252.\\
Delfosse, X. \etal (1998) Rotation and chromospheric activity in field M dwarfs. \textit{Astron. \& Astrophys.} 331, 581-595.\\
Deming, D.(2007) Spitzer Transit and Secondary Eclipse Photometry of GJ 436b. \textit{Astrophys. J.} 667, L199-L202.\\
Dickey, J.O. \etal (1994) Lunar Laser Ranging - a Continuing Legacy of the Apollo Program. \textit{Science} 265, 482-490.\\
Eggleton, P.P \etal (1998) The Equilibrium Tide Model for Tidal Friction. \textit{Astrophysical J.} 499, 853-870.\\
Fortney, J.J., Marley, M.S. \& Barnes, J.W. (2007) Planetary Radii across Five Orders of Magnitude in Mass and Stellar Insolation: Application to Transits. \textit{Astrophys. J.} 659, 1661-1672.\\
Franck, S. \etal (2000) Determination of habitable zones in extrasolar planetary systems: Where are Gaia's sisters? \textit{J. of Geophys. Research} 105, 1651-1658.\\
Greenberg, R. (1977) Orbit-orbit resonances among natural satellites. In \textit{Planetary Satellites}, edited by J. Burns. Arizona UP, Tucson, AZ.\\
Gillon, M. \etal (2007) Accurate Spitzer infrared radius measurement for the hot Neptune GJ 436b. \textit{Astron. \& Astrophys.} 471, L51-L54.\\
Goldreich, P. (1966) Final spin states of planets and satellites. \textit{Icarus} 71, 1-7.\\
Goldreich, P. \& Soter, S. (1966) Q in the Solar System. \textit{Icarus} 5, 375-389.\\
Hut, P. (1981) Tidal evolution in close binary systems. \textit{Astron. \& Astrophys.} 99, 126-140.\\
Jackson, B., Greenberg, R. \& Barnes, R. (2007) Tidal Evolution of Close-in Extrasolar Planets. \textit{Astrophys. J.}, submitted.\\
Kaltenegger, L., Traub, W.A., \& Jucks, K.W. (2007) Spectral Evolution of an Earth-like Planet. \textit{Astrophys. J.} 658, 598-616.\\
Kasting, J.F. \& Catling, D. (2003) Evolution of a Habitable Planet. \textit{Ann. Rev. of Astron. \& Astrophys.} 41, 429-463.\\
Kasting, J.F., Whitmire, D.P., \& Reynolds, R.T. (1993) Habitable Zones around Main Sequence Stars. \textit{Icarus} 101, 108-128.\\
Kaula, W.M. (1964) Tidal Dissipation by Solid Friction and the Resulting Orbital Evolution. \textit{Rev. Geophysics} 2, 661-685.\\
Kiang, N.Y. \etal (2007) Spectral Signatures of Photosynthesis. II. Coevolution with Other Stars And The Atmosphere on Extrasolar Worlds. \textit{Astrobiology} 7, 252-274.\\
Lambeck, K. (1977) Tidal Dissipation in the Oceans: Astronomical, Geophysical and Oceanographic Consequences. \textit{Phil. Trans. of the Royal Soc. of London A} 287, 545-594.\\
Laughlin, G., Bodenheimer, P. \& Adams, F.C. (2004) The Core Accretion Model Predicts Few Jovian-Mass Planets Orbiting Red Dwarfs. \textit{Astrophys. J. Lett.} 612, L73-L76.\\
Lee, M.-H. \& Peale, S.J. (2002) Dynamics and Origin of the 2:1 Orbital Resonances of the GJ 876 Planets. \textit{Astrophys. J.} 567, 596-609.\\
L\'eger, A. \etal (2004) A new family of planets? ``Ocean-Planets''. \textit{Icarus} 169, 499-504.\\
Lissauer, J.J. (2007)  Planets Formed in Habitable Zones of M Dwarf Stars Probably Are Deficient in Volatiles. \textit{Astrophys. J. Lett.} 660, L159-L162.\\
Lovelock, J.E. \& Margulis, L. (1974) Homeostatic tendencies of the Earth's atmosphere. \textit{Origins of Life and Evolutions of Biospheres} 5, 93-103.\\
Mardling, R.A (2007) Long-term tidal evolution of short-period planets with companions. \textit{MNRAS}, accepted.\\
Mardling, R.A. \& Lin, D.N.C. (2002)  Calculating the Tidal, Spin, and Dynamical Evolution of Extrasolar Planetary Systems. \textit{Astrophys. J.} 573, 829-844.\\
-----------. (2004) On the Survival of Short-Period Terrestrial Planets. \textit{Astrophysical J.} 614, 955-959.\\
Naef, D. \etal (2007) The HARPS search for southern extra-solar planets. IX. Exoplanets orbiting HD 100777, HD 190647, and HD 221287. \textit{Astron. \& Astrophys.} 470, 721-726.
Ogilvie, G.I. \& Lin, D.N.C. (2007) Tidal Dissipation in Rotating Solar-Type Stars. \textit{Astrophys. J.} 661, 1180-1191.\\
Peale, S.J. (1977) Rotation Histories of the Natural satellites. In \textit{Planetary Satellites}, edited by J. Burns. Arizona UP, Tucson, AZ.\\
Peale, S.J., Cassen, P, \& Reynolds ,R.T. (1979) Melting of Io by tidal dissipation. \textit{Science} 203, 892-894.\\
Rasio, F.A. \etal (1996) Tidal Decay of Close Planetary Orbits. \textit{Astrophys. J.} 470, 1187-1191.\\
Raymond, S.N., Quinn, T. \& Lunine, J.I. (2004) Making other earths: dynamical simulations of terrestrial planet formation and water delivery. \textit{Icarus} 168, 1-17.\\ 
------------. (2007) High-Resolution Simulations of The Final Assembly of Earth-Like Planets. 2. Water Delivery And Planetary Habitability. \textit{Astrobiology} 7, 66-84.\\
Raymond, S.N., Scalo, J. \& Meadows, V. (2007) A decreased probability of habitable planet formation around low-mass stars. \textit{Astrophys. J.} 669, 606-614.\\
Regenauer-Lieb, K., Yuen, D.A. \& Barlund, J. (2001) The Initiation of Subduction: Criticality by Addition of Water? \textit{Science} 294, 578-580.\\
Ribas, I., private communication.\\
Sagan, C. \etal (1993) A Search for Life on Earth from the Galileo Spacecraft. \textit{Nature} 365, 715-721.\\
Scalo, J. \etal (2007) M Stars as Targets for Terrestrial Exoplanet Searches And Biosignature Detection. \textit{Astrobiology} 7, 85-166.\\
Selsis, F., private communication.\\
Selsis, F. \etal (2007) Habitable planets around the star Gl581? \textit{Astron. \& Astrophys.}, accepted.\\
Segura, A. \etal (2005) Biosignatures from Earth-Like Planets Around M Dwarfs. \textit{Astrobiology} 5, 706-725.\\
Sotin, C., Grasset, O., \& Moquet, A. (2007) Mass/radius for extrasolar Earth-like planets and ocean planets. \textit{Icarus}, in press.\\
Tarter, J.C \etal (2007) A Reappraisal of The Habitability of Planets around M Dwarf Stars. \textit{Astrobiology} 7, 30-65.\\ 
Tinetti, G., Rashby, S. \& Yung, Y.L. (2006) Detectability of Red-Edge-shifted Vegetation on Terrestrial Planets Orbiting M Stars. \textit{Astrophys. J. Lett.} 644, L129-L132.\\
Tinetti, G. \etal (2006a) Detectability of Planetary Characteristics in Disk-Averaged Spectra. I: The Earth Model. \textit{Astrobiology} 6, 34-47.\\
-----------. (2006b) Detectability of Planetary Characteristics in Disk-Averaged Spectra II: Synthetic Spectra and Light-Curves of Earth. \textit{Astrobiology} 6, 881-900.\\
von Bloh, W. \etal (2007) The habitability of super-Earths in Gliese 581. astro-ph/07053758.\\
Udry, S. \etal (2007) The HARPS search for southern extra-solar planets XI. Super-Earths (5 \& 8 $M_\oplus$) in a 3-planet system. \textit{Astron. \& Astrophys. Lett.} 469, L43-L47.\\
Valencia, D., Sasselov, D.D., \& O'Connell, R.J. (2007) Radius and Structure Models of the First Super-Earth Planet. \textit{Astrophys. J.} 656, 545-551.\\
Walker, J.C.G., Hays, P.B. \& Kasting, J.F. (1981) A negative feedback mechanism for the long-term stabilization of Earthy surface temperature. \textit{J. Geophys. Res.} 86, 9776-9782.\\
Williams, D.M., Kasting, J.E. \& Wade, R.A. (1997) Habitable moons around extrasolar giant planets. \textit{Nature} 385, 234-236.\\
Williams, D.M. \& Pollard, D. (2002) Earth-like worlds on eccentric orbits: excursions beyond the habitable zone. \textit{Int. J. of Astrobiology} 1, 61-69.\\
Yoder, C.F. (1995) Astrometric and geodetic properties of Earth and the solar system. In \textit{Global Earth Physics. A Handbook of Physical Constants}, edited by T. Ahrens, American Geophysical Union, Washington, D.C.\\

\clearpage
\begin{center}Table 1: Best-Fit Orbital Elements of the Gl 581 Planetary System
\end{center}
\begin{tabular}{ccccccc}
\hline
Planet & $m$ (M$_{\oplus}$) & $P$ (d) & a (AU) & $e$ & $\varpi$ ($^o$) & $T_{peri}$ (JD)\\
\hline\hline
b & 15.7 & 5.363 & 0.041 & 0.02 & 273 & 2452998.76\\
c & 5.03 & 12.932 & 0.073 & 0.16 & 267 & 2452993.38\\
d & 7.7 & 83.6 & 0.25 & 0.2 & 295 & 2452936.9\\
\end{tabular}

\clearpage
Figure 1 -- Semi-major axes that are habitable as a function of
stellar mass and eccentricity from Eqs.\ (\ref{eq:lin} --
\ref{eq:luminosity}). The shading represents
different choices of cloud cover for the limits of the HZ. Darkest
gray assumes no cloud cover, medium gray assumes 50\% cloud cover,
light gray assumes total cloud cover, and white is uninhabitable.

Figure 2 -- Contours of $\tau_{HL}$ of terrestrial planets, that begin
at the inner edge of the 0\% cloud cover EHZ, as a function of stellar
mass and initial eccentricity $e_0$. The terrestrial planets have a mass of
5 $M_\oplus$ (top left), 1 $M_\oplus$ (top right), 0.5 $M_\oplus$
(bottom left), and 0.1 $M_\oplus$ (bottom right), and a radius
determined by Eq.\ (\ref{eq:rp}). Planets with initial eccentricities
below 0.53, or orbiting stars with a mass greater then 0.35 $M_\odot$
are habitable for at least 10 Gyr. Lower mass planets tend to have
longer habitable lifetimes.\\

\begin{figure}
\plotone{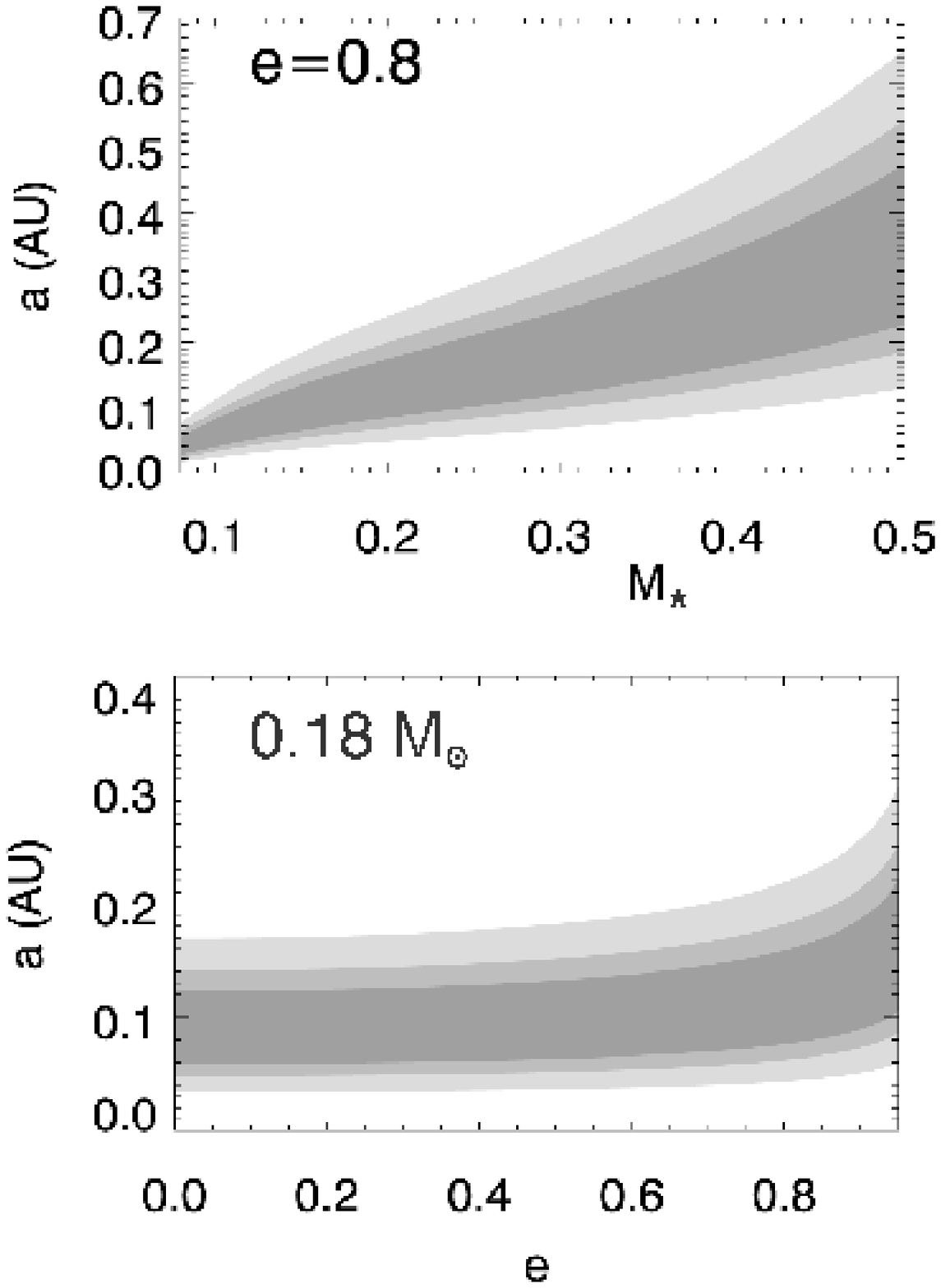}
\figcaption[]{\label{fig:hz} \small{}}
\end{figure}

\begin{figure}
\plotone{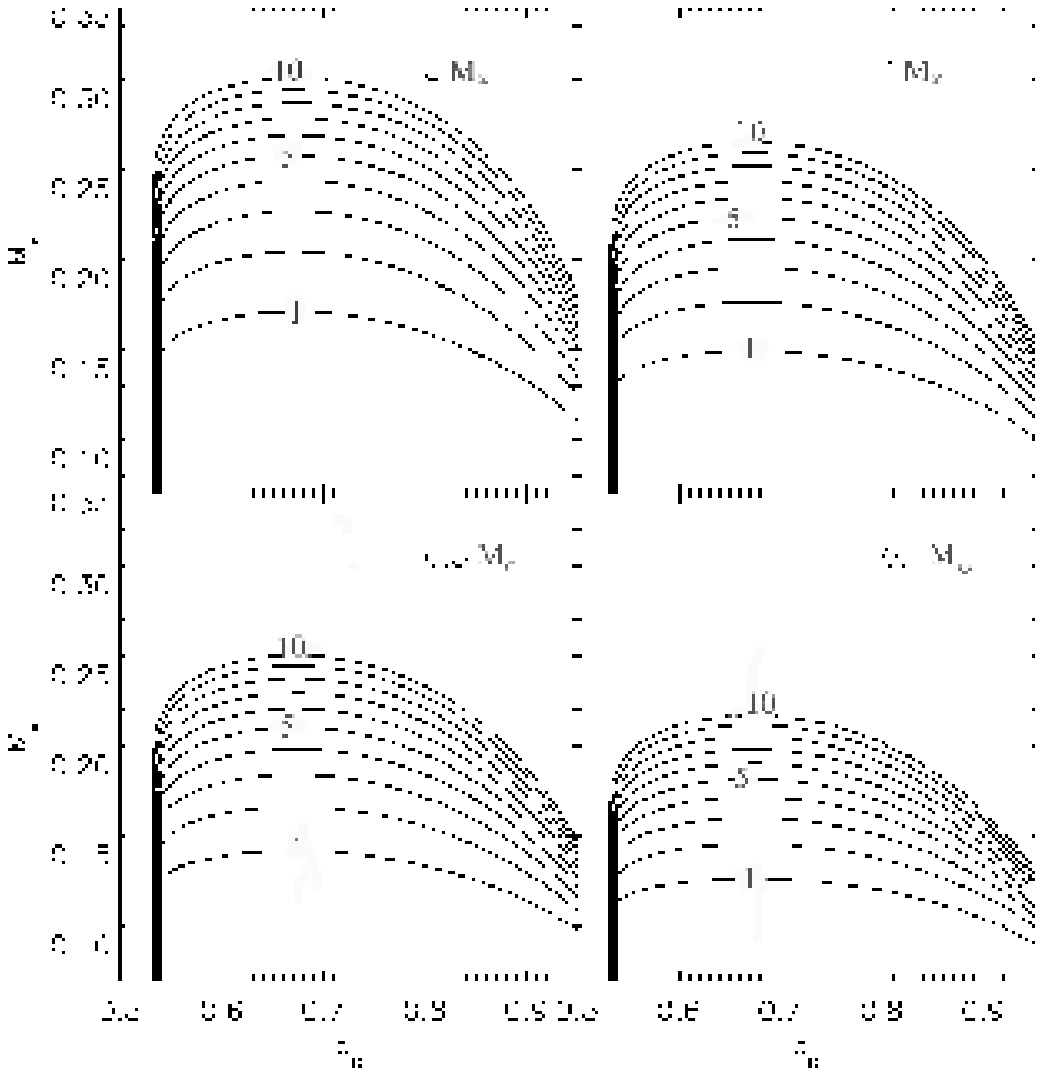}
\figcaption[]{\label{fig:thz4} \small{}}
\end{figure}

\begin{figure}
\plotone{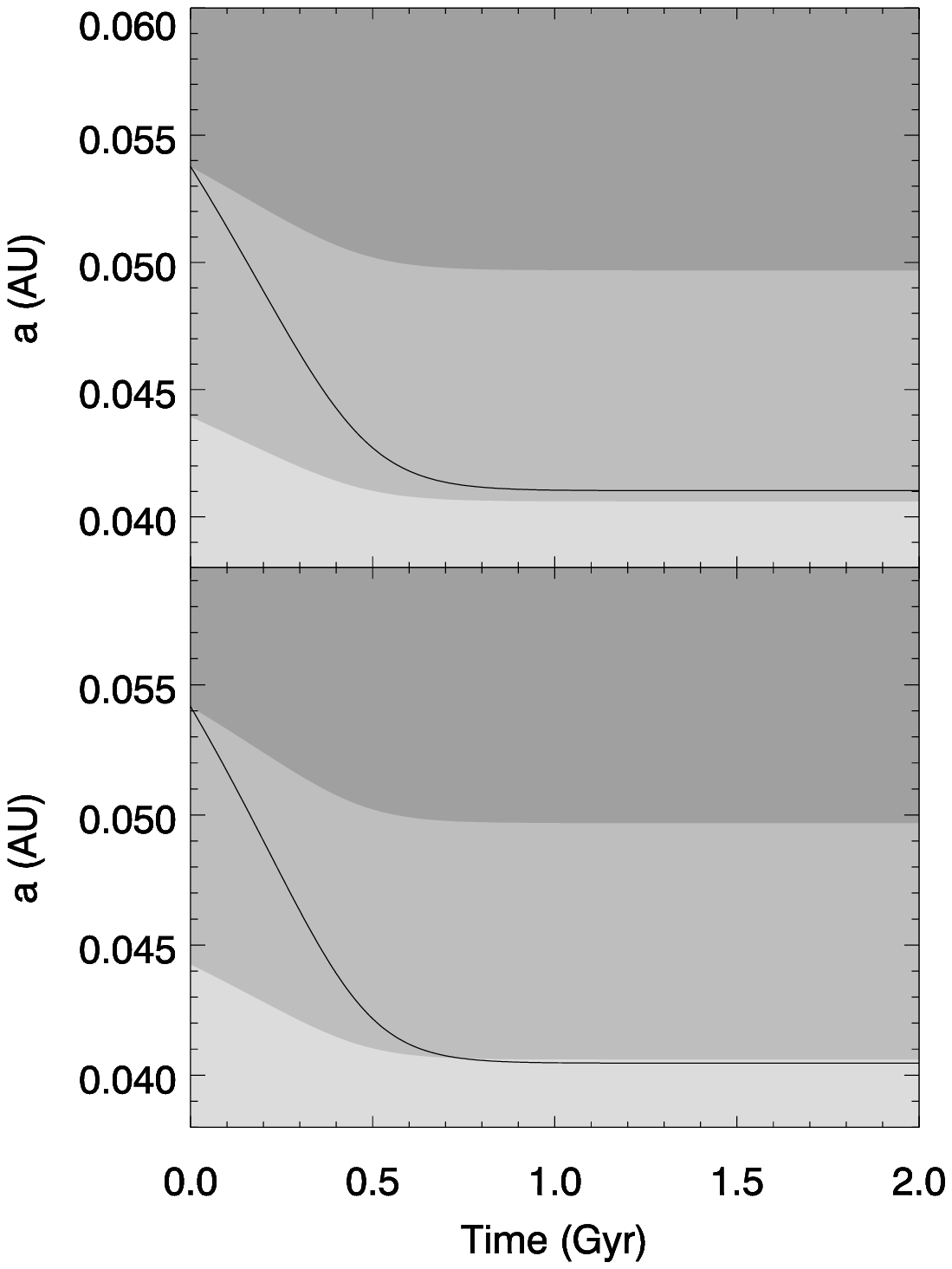}
\figcaption[]{\label{fig:evol} \small{}}
\end{figure}

\begin{figure}
\plotone{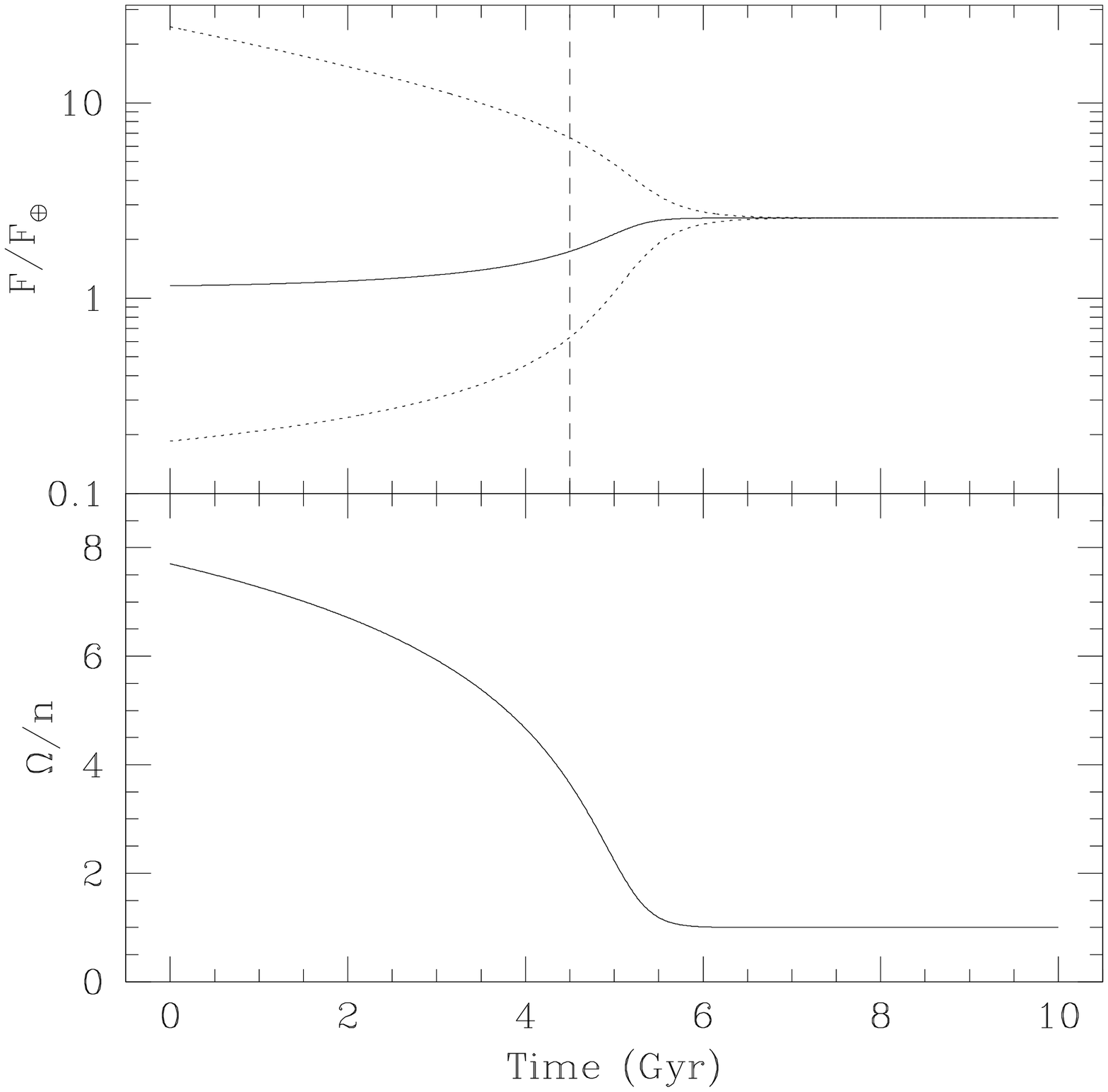}
\figcaption[]{\label{fig:flux} \small{}}
\end{figure}

\begin{figure}
\plotone{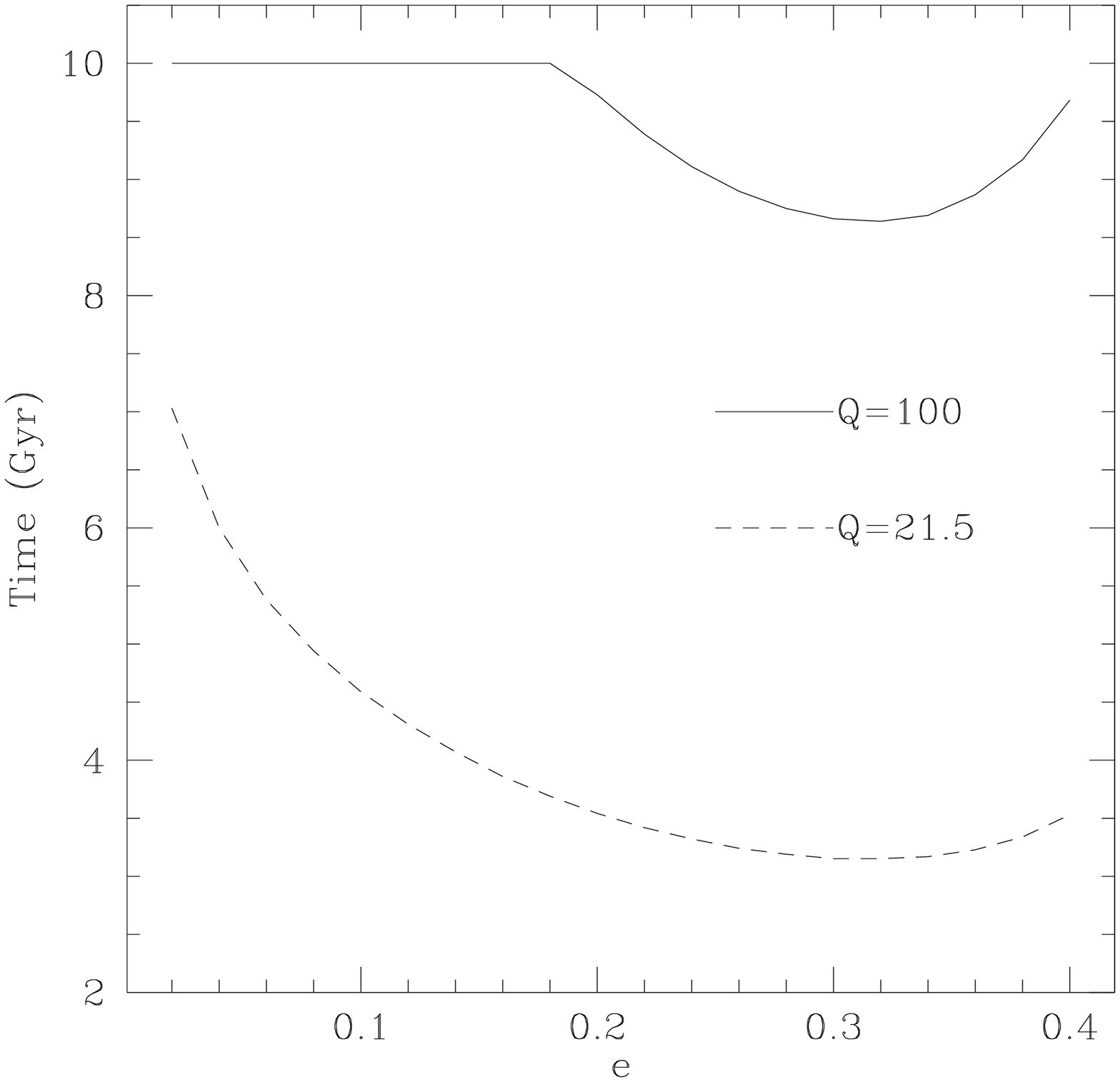}
\figcaption[]{\label{fig:e} \small{}}
\end{figure}

\begin{figure}
\plotone{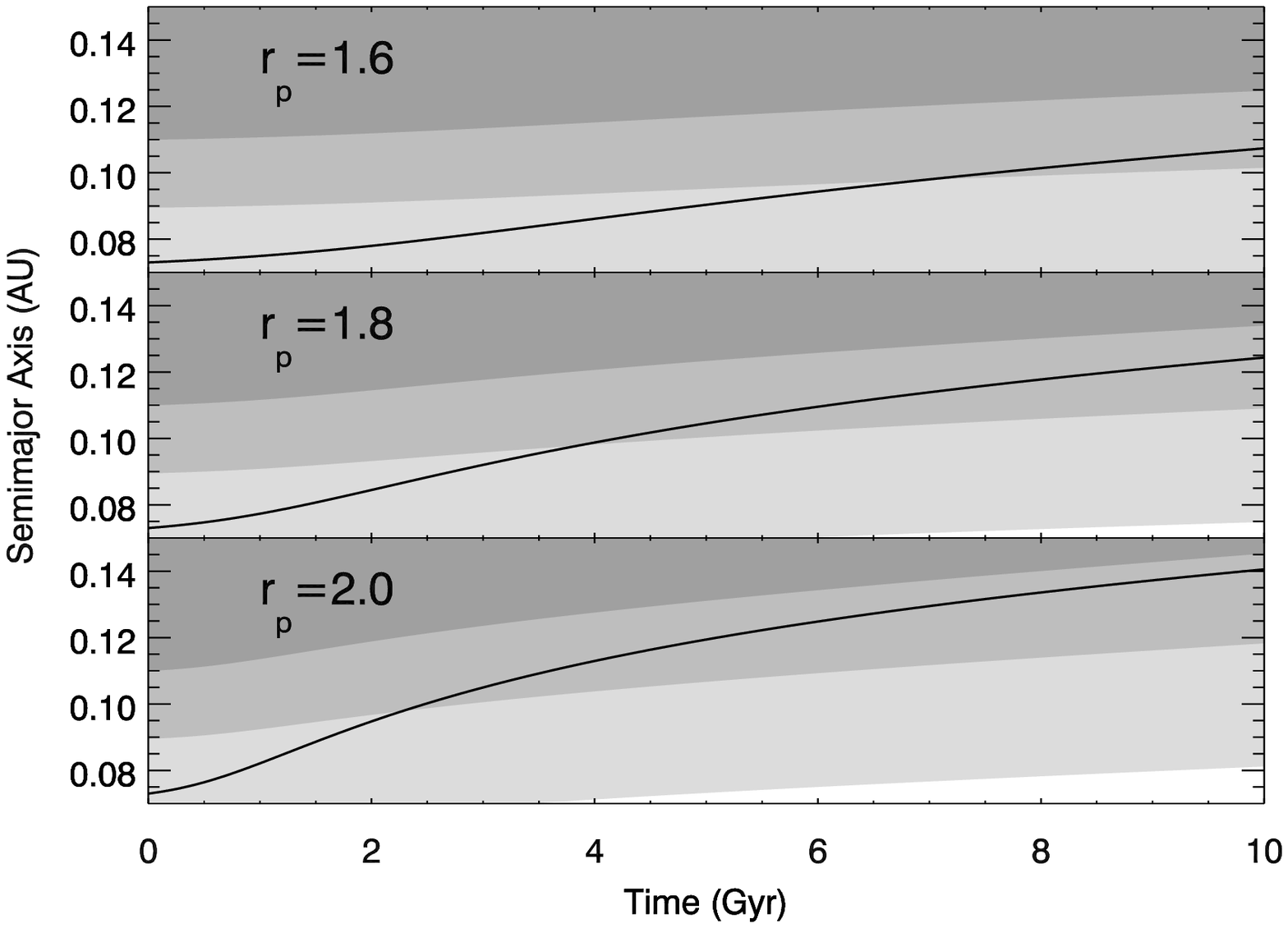}
\figcaption[]{\label{fig:rp} \small{}}
\end{figure}

\begin{figure}
\plotone{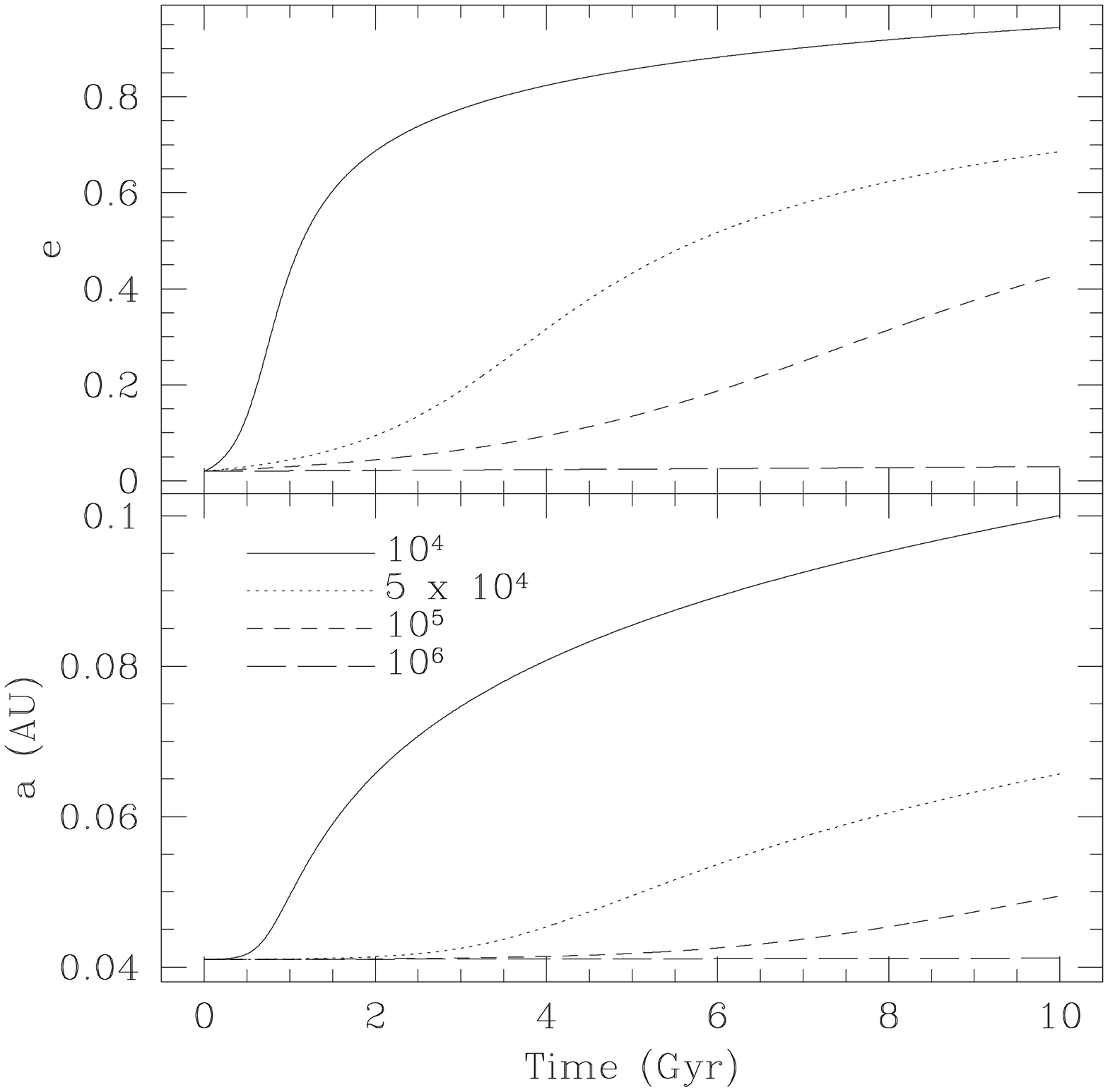}
\figcaption[]{\label{fig:qb} \small{}}
\end{figure}

\begin{figure}
\plotone{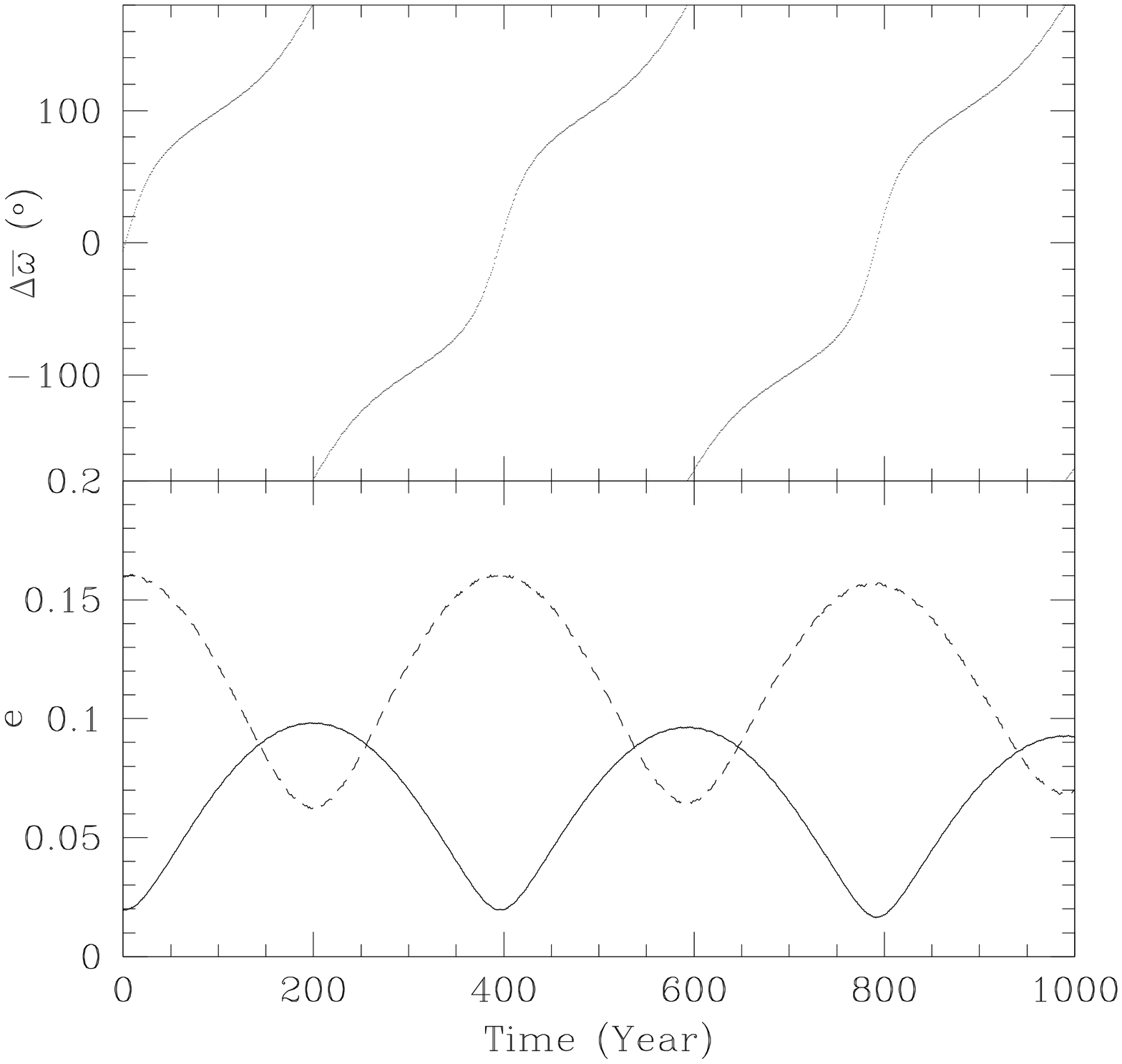}
\figcaption[]{\label{fig:secular} \small{}}
\end{figure}

\begin{figure}
\plotone{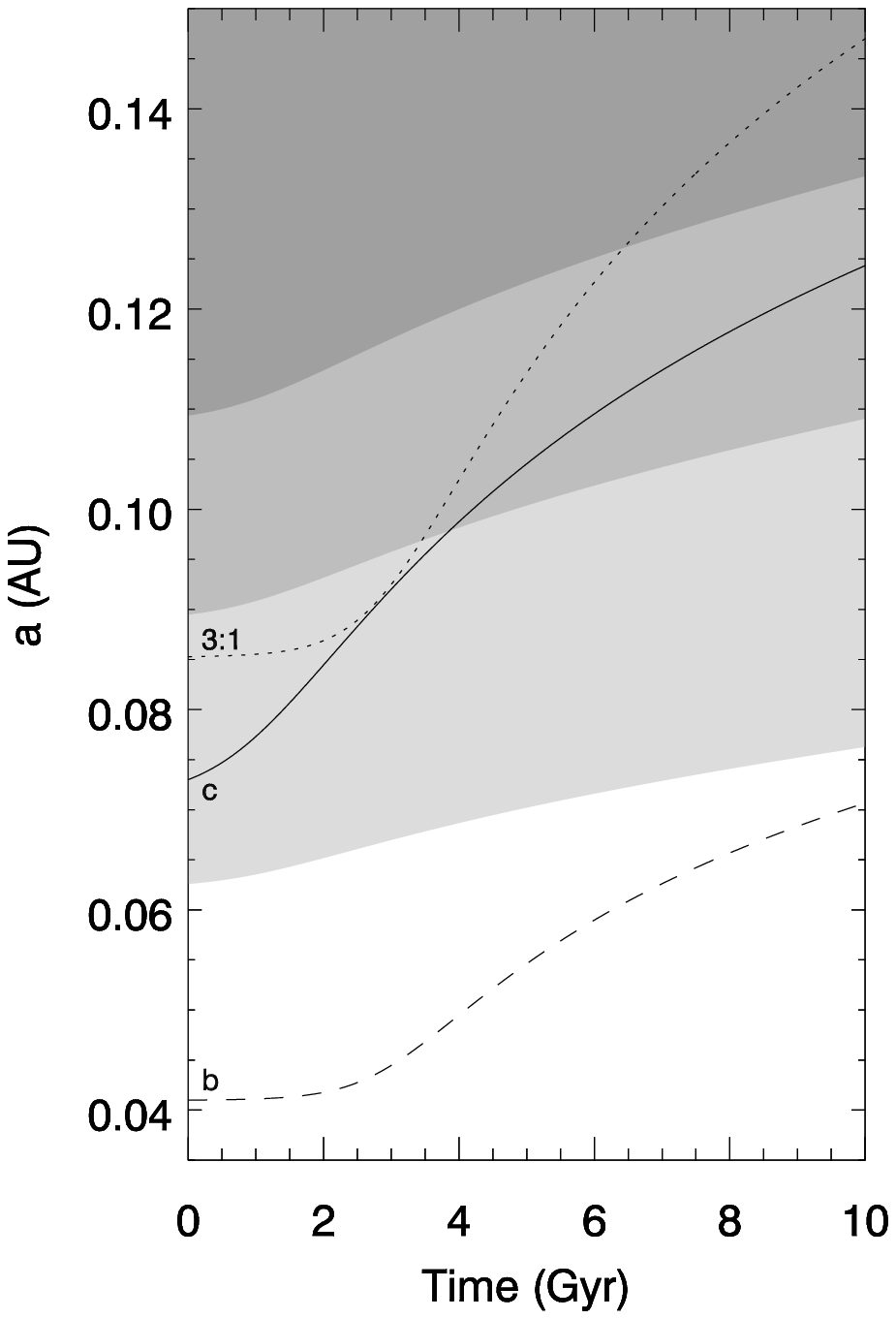}
\figcaption[]{\label{fig:3_1} \small{}}
\end{figure}

\begin{figure}
\plotone{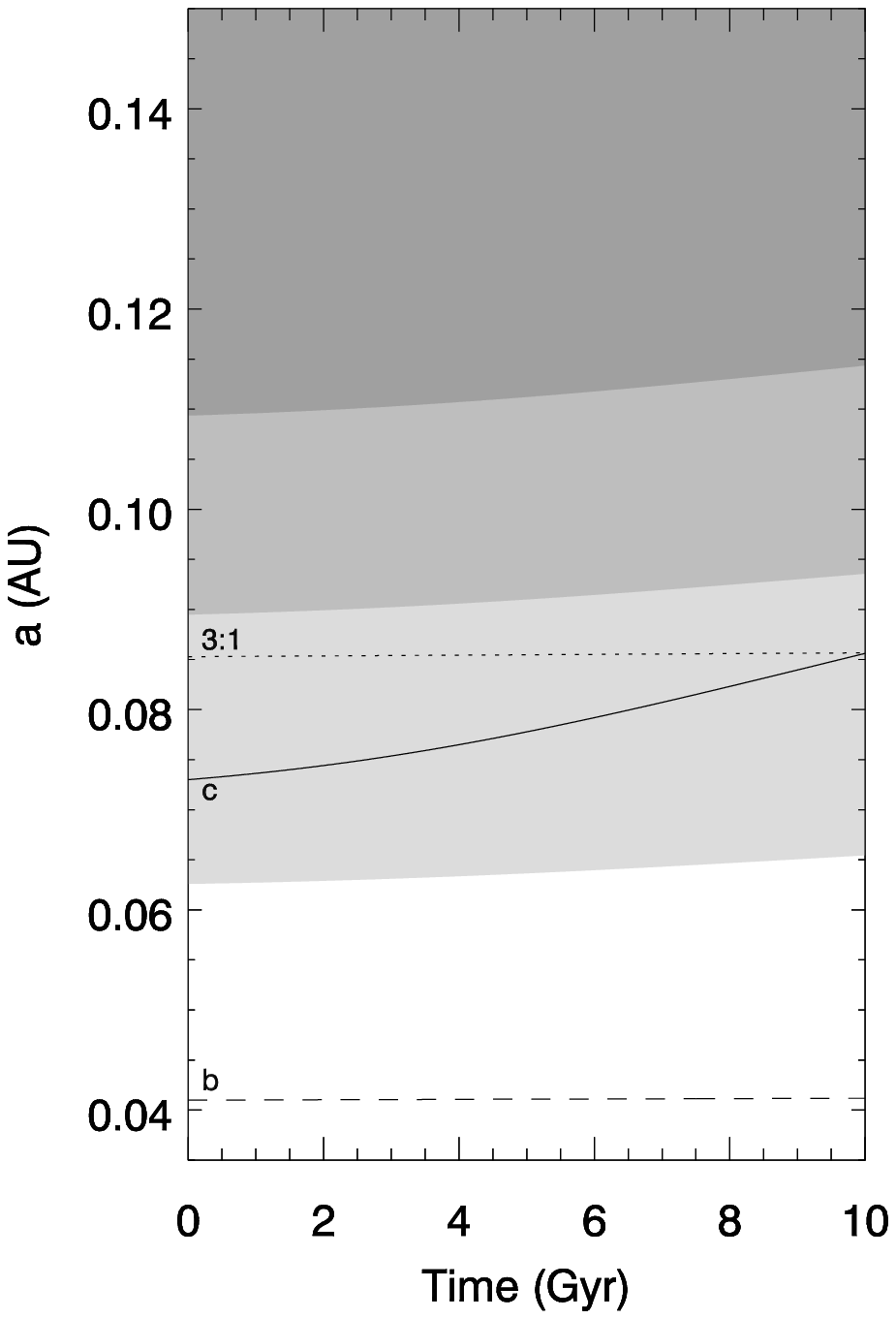}
\figcaption[]{\label{fig:bestrewind} \small{}}
\end{figure}

Figure 3 -- Evolution of the semi-major axes (solid line) for two hypothetical planets relative to EHZ boundaries (shown by the shading, \cf Fig.\ \ref{fig:hz}). These planets
are in orbit about a 0.15 $M_\odot$ star. \textit{Top:} Evolution of a
planet with $e_0 = 0.52$. This planet's semi-major axis decreases
until its eccentricity reaches zero after 750 million years, at which
point the tidal evolution effectively stops. The tidal evolution
stopped just before the planet crossed the inner edge of the EHZ, and
the planet will be habitable indefinitely. \textit{Bottom:}
Evolution of a planet with $e_0 = 0.54$. Although very similar to the
other planet's evolution, this planet's larger initial eccentricity
results in more torque so that when $e$ reaches zero, the planet is
just interior to the 50\% cloud cover HZ. The habitable lifetime for this planet is $7.5 \times 10^8$ years.\\

Figure 4 -- \textit{Top:} Orbit-averaged stellar flux
received by a tidally evolving planet relative to that of the Earth
$F_\oplus$ (solid line), see Eq.\ (\ref{eq:flux}), as well as at periastron and
apoastron (dotted lines). The planet orbits a 0.2 M$_\oplus$ star with
initial orbital elements of $a_0 = 0.0877$, and $e_0 = 0.84$. The tidal
lifetime of the planet, 4.5 Gyr in this case, corresponds to the age of
the Earth, and is shown by the dashed line. Note, however, the extreme
difference between flux at periastron and apoastron (a factor of
200). \textit{Bottom:} Ratio of the rotation frequency of the tidally
evolving planet to its instantaneous mean motion, \cf Eq.\
(\ref{eq:spinlock}). Initially the planet rotates nearly 10 times faster
than it revolves, but after $\sim 6$ Gyr, the orbit has circularized and
$\Omega = n$.\\

Figure 5 -- Time since Gl 581 c would have been habitable
as a function of its current eccentricity. We set $R_c = 1.8
R_\oplus$, and consider $Q$ values of 100 and 21.5. The $Q = 21.5$
case permits habitability as recently as 3 Gyr ago.\\

Figure 6 -- Evolution of Gl 581 c for three different possible radius
values. The shading represents different definitions of the HZ, \cf Fig.\ \ref{fig:hz}.\\

Figure 7 -- Evolution of $a_b$ and $e_b$ for various choices of
$Q_b$, assuming $k = 3/2$. The evolution is negligible unless $Q_b \lsim 10^5$.\\

Figure 8 -- Apsidal evolution of the Gl 581 b and c planets, see Table
1. \textit{Top:} The apsidal oscillation is
circulation. \textit{Bottom:} Evolution of $e_b$ (solid line) and
$e_c$ (dashed line). The orbits are currently near the extrema of
their eccentricity values.\\

Figure 9 -- Evolution of planets b (dashed line), c (solid line) and
the instantaneous location of b's outer 3:1 mean motion resonance
(dotted line) for $Q_b = 4\times 10^4, k = 3/2$. In this case planet c would have
just avoided the 3:1 resonance. The shading corresponds to different definitions of the HZ, \cf Fig.\ \ref{fig:hz}. 10 Gyr ago the configuration was
unstable.\\

Figure 10 -- Evolution of planets b (short-dashed line), c (solid line) and the instantaneous location of b's outer 3:1 mean motion resonance (dotted line) for $Q_b = 10^6, R_b = 3.74 R_\oplus, k = 3/2$ and $Q_c = 100, R_c = 1.8 R_\oplus, k = 0.3$. In our model, this evolution is the most believable, and prevents planet c from crossing the 3:1 mean motion resonance and reaching the 50\% cloud cover habitable zone (medium grey). This evolution permits stable interactions 10 Gyr ago.

\clearpage

\clearpage

\clearpage

\clearpage

\clearpage

\clearpage

\clearpage

\clearpage

\clearpage

\clearpage

\end{document}